\documentstyle[prb,aps,eqsecnum,floats,epsf]{revtex}

\def\mathrm{\rm}

\makeatletter
\def\preprint#1{%
\def\@preprint{\noindent\hfill\hbox{#1}\vskip 10pt}%
}
\makeatother

\begin{document}

\draft
\preprint{OUTP-97-10S, ITP-UH-06/97}

\title{X-ray edge singularity in integrable lattice models\\
         of correlated electrons}
\vspace{1.5 em}

\author{Fabian H.\,L.\,E\ss{}ler$^{(a)}$ and Holger Frahm$^{(b)}$}
\address{$^{(a)}$Department of Physics, Theoretical Physics,
        Oxford University\\ 1 Keble Road, Oxford OX1 3NP, United Kingdom}
\address{$^{(b)}$Institut f\"ur Theoretische Physik,
  Universit\"at Hannover, D-30167~Hannover, Germany}
\date{\today}
\maketitle
\begin{abstract}
  We study the singularities in X-ray absorption spectra of
  one-dimensional Hubbard and $t$--$J$ models. We use Boundary
  Conformal Field Theory and the Bethe Ansatz solutions of
  these models with both periodic and open boundary conditions 
  to calculate the exponents describing the power-law decay near 
  the edges of X-ray absorption spectra in the case where the
  core-hole potential has bound states.
\end{abstract}
%

%
%
\section{Introduction}
\begin{figure}[ht]
\begin{center}
\noindent
\epsfxsize=0.9\textwidth
\epsfbox{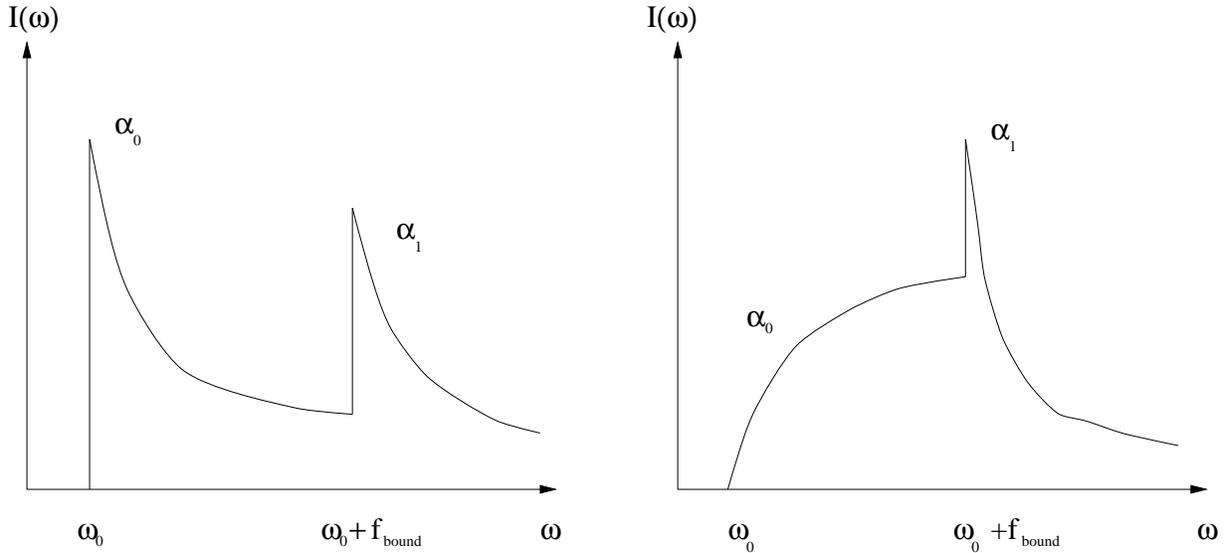}
\end{center}
\caption{\label{fig:int}%
X-ray absorption rate as a function of frequency: a) $\alpha_0>0$,
$\alpha_1>0$\ , b) $\alpha_0<0$, $\alpha_1>0$,}
\end{figure}

X-ray absorption in a metal can be described by a simple model put forward by
Nozi{\`e}res and de~Dominicis\cite{nodo:69}. An electron from a filled inner
shell of one of the nuclei is raised into the conduction band. This generates
a local potential $V$ at the position of the nucleus that lost the
core-electron, which in turn acts on the (noninteracting) conduction-band
electrons and affects the X-ray absorption probability. The situation is
described by the Hamiltonian
\begin{equation}
  H = \sum_{\vec{k}} \epsilon({\vec{k}})\ c^\dagger({\vec{k}})c({\vec{k}})
    + bb^\dagger \sum_{{\vec{k}},{\vec{k}}^\prime}
    V({\vec{k}},{\vec{k}}^\prime)\
    c^\dagger({\vec{k}})c({{\vec{k}}^\prime})+ E_0 b^\dagger b\ , 
\label{hfermi}
\end{equation}
where $\epsilon({\vec{k}})$ is the dispersion of the conduction band
electrons, $b^\dagger$ and $b$ ($c({\vec{k}})$ and
$c^\dagger({\vec{k}})$) are annihilation and creation operators for
the core-hole (for conduction band electrons with wavevector
${\vec{k}}$) and $E_0$ is the energy of the core state. As $b^\dagger
b$ commutes with $H$, the Hilbert space splits into two sectors: in
one the core-level is filled ($bb^\dagger=0$) and there is no
potential whereas in the other one the core level is empty ($bb^\dagger
=1$) and $V$ acts on the conduction electrons. As was shown in
Ref.~\onlinecite{nodo:69} the inner core disturbance acts as a
transient one-body potential on the conduction electrons, which means
that one needs to study the response of the conduction band electrons
to the potential $V$ applied between times $t=0$ and $t=t^\prime$. The
X-ray absorption rate can be expressed by the golden rule as
\begin{equation}
   I(\omega)\propto \sum_n|\langle n|c^\dagger_0(0)|0\rangle|^2\ 
   \delta(\omega+E_{GS}-E_n-E_0)\ ,
\label{io}
\end{equation}
where $c_0(t)$ annihilates a conduction-band electron at position
$\vec{x}=0$ at time $t$, $|0\rangle$ is the ground state at times $t<0$ and
$H|0\rangle=E_{GS}|0\rangle$. The r.h.s.\ of (\ref{io}) can be expressed
in terms of the spectral representation of the Fourier transform of the
retarded correlation function $\langle\!\langle b^\dagger(t)c_0(t)
c^\dagger_0(0) b(0)\rangle\!\rangle$ so that
\begin{equation}
  I(\omega)\propto\Im m\int_0^\infty\ dt\ e^{i\omega t} 
    \langle\!\langle b^\dagger(t)c_0(t)
                     c^\dagger_0(0) b(0)\rangle\!\rangle\ .
\end{equation}
Near the threshold $\omega_0\approx E_0$ the intensity $I(\omega)$ displays
a characteristic singularity of the form
\begin{equation}
  I(\omega) \sim {1\over |\omega-\omega_0|^\alpha}\ .
\label{powerlaw}
\end{equation}
For the system (\ref{hfermi}) the critical exponent $\alpha$ has been
determined exactly and is expressed in terms of the phase shift at the Fermi
surface \cite{nodo:69,scho:69}.
A very interesting case is the one where the local potential $V$ is
sufficiently strong to bind a conduction electron \cite{cono} (see also
Refs.~\onlinecite{abrx:72,affl:96}). In this case the absorption
spectrum (if $\alpha>0$) features two thresholds with characteristic power-law
decays of $I$ as a function of $\omega$ (see Fig.~\ref{fig:int}a). If
$\alpha<0$ there is no discontinuity and $I(\omega)$ goes to zero instead (see
Fig.~\ref{fig:int}b).

In the present work we wish to investigate the analogous situation for
integrable lattice models of strongly interacting conduction electrons in one
dimension\cite{vladb,koes:rep}. These models are particular realizations of
{\sl Luttinger liquids} and the X-ray problem for such systems has been
investigated by various authors (a detailed pedagocial discussion can be found
in the forthcoming book\cite{alexeibook}). The case of a core potential with
no backscattering was solved in Refs.~\onlinecite{NoBS} and the case of a
perfectly reflecting potential was treated in Ref.~\onlinecite{prok:94}. The
general case was investigated by Affleck and Ludwig \cite{aflu:94} using
Boundary Conformal Field Theory (BCFT) \cite{BCFT}. Recently,
Affleck \cite{affl:96} reconsidered the X-ray problem for a Fermi liquid
(\ref{hfermi}) for the case where $V$ has a bound state from the point of view
of BCFT. This motivated the present work in which we study the X-ray problem
in Hubbard and $t$-$J$ chains for core hole potentials with bound states. Let
us discuss the general setup for the case of the Hubbard model.  At times
$t<0$ we take the system to be periodic
\begin{equation}
H_A  = - \sum_{j=1}^{L} \sum_\sigma\left(c^\dagger_{j,\sigma}c_{j+1,\sigma}
       +c^\dagger_{j+1,\sigma}c_{j,\sigma}\right)
       + 4u\sum_{j=1}^L n_{j\uparrow}n_{j\downarrow}
       + \mu \hat{N} .
\end{equation}
At time $t=0$ we switch on the core potential $V_{1L}$ acting on sites $1$ and
$L$ (a similar situation has been studied in \cite{pesc:84}). In the
general case this potential will include a backscattering term which
will then drive the system to the open chain fixed point \cite{kane},
{\sl i.e.} break the chain across the link $1L$. We model this
situation by considering the Hamiltonian 
\begin{equation}
H_B = -\sum_{j=1}^{L-1} \sum_\sigma\left(c^\dagger_{j,\sigma}c_{j+1,\sigma}
       +c^\dagger_{j+1,\sigma}c_{j,\sigma}\right)
       +4u\sum_{j=1}^L n_{j\uparrow}n_{j\downarrow}
       +\mu \hat{N} +H_1+H_L\ ,
\end{equation}
where $H_{1,L}$ are one-body interactions acting on sites $1$ and $L$
respectively. At time $t$ we switch off the core potential which changes
the Hamiltonian back to $H_A$.  Depending on the precise form of the
interactions $H_{1,L}$ bound states can be formed at the boundaries. 
As the elementary excitations in the Hubbard model are not electrons
like in the case of the Fermi liquid discussed above but (anti)holons
and spinons one has to consider several possibilities:  In addition to
the case in which there are no bound states the core-hole potential
can bind either a spinon, a (anti)holon, both a spinon and a
(anti)holon or, for an attractive boundary potential of the order of
the Hubbard interaction $4u$, a pair of electrons. 

\begin{figure}[ht]
\begin{center}
\noindent
\epsfxsize=0.6\textwidth
\epsfbox{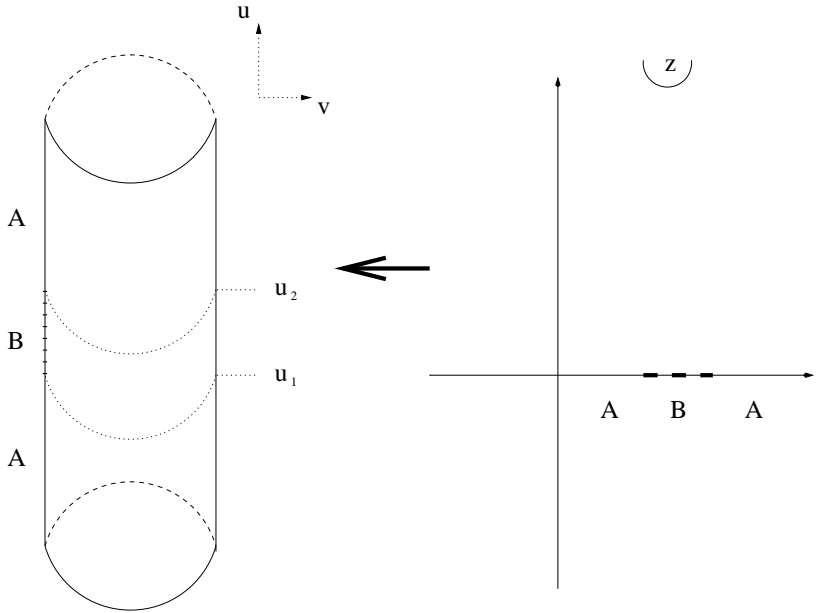}
\end{center}
\caption{\label{fig:cyl}%
The mapping from the cylinder to the plane.}
\end{figure}

In order to extract the X-ray exponent we use BCFT and the fact that the
low-energy spectrum of both Hubbard and $t$-$J$ models can be described in
terms of two $c=1$ Conformal Field Theories or equivalently a spin and charge
separated Luttinger liquid \cite{frko:90,kaya:91}.  Our discussion closely
follows Ref.~\onlinecite{aflu:94}.
We start by considering the Luttinger liquid defined on the complex plane
with coordinate $z$.  Identifying the radial part of $z$ with the time
variable the case of periodic boundary conditions (A) is realized if we
consider the complex plane without boundaries.  The change to open boundary
conditions (B) corresponds to the introduction of a cut in the plane from
$z_0$ to $z_1$.
As explained above this change of boundary conditions corresponds to
switching on (and off) the core-hole potential.  Choosing $0<\tau_0=z_0
<\tau_1=z_1$ real and mapping the plane to a cylinder {\sl via} the
conformal transformation $z=\exp({2\pi(u+ i v)}/{L})$ this cut gets mapped
onto a seam in the time direction of the cylinder (see Fig.~\ref{fig:cyl}).

The Green's function of an operator ${\cal O}$ with dimension $x$ on the
complex plane without boundaries is given by
\begin{equation}
   \langle A|{\cal O}(\tau_1){\cal O}^\dagger(\tau_2)|A\rangle =
   \frac{1}{(\tau_1-\tau_2)^{2x}}\ .
\label{greenc}
\end{equation}
The Green's function on the cylinder is obtained by the conformal mapping.
For $u_2-u_1\gg L$ we obtain
\begin{equation}
   \langle A|{\cal O}(u_1){\cal O}^\dagger(u_2)|A\rangle \sim
   \left(\frac{2\pi}{L}\right)^{2x}\
   e^{-\frac{2\pi x}{L}(u_2-u_1)}\ .
\end{equation}
To study the edge singularity we choose ${\cal O}^\dagger$ to be an
operator which changes the boundary conditions from A to B.  The
same correlation function can be evaluated alternatively by inserting a
resolution of the identity in terms of the eigenstates $|B;\nu\rangle$ of
the system with reflecting boundary conditions
\begin{equation}
  \langle A|{\cal O}(u_1){\cal O}^\dagger(u_2)|A\rangle =\sum_\nu
  |\langle A|{\cal O}(0)|B;\nu\rangle|^2\ e^{-(E_B^\nu-E_A)(u_2-u_1)}\ .
\end{equation}
The leading contribution to this sum comes from the ground state or a low
lying excited state (this depends on the operator ${\cal O}$ because the form
factor must be nonvanishing) with boundary condition of type B.  Comparing
the two expressions for the correlation functions on the cylinder
allows one to extract the scaling dimensions of the boundary changing
operator ${\cal O}$ 
\begin{equation}
   x_\nu=\frac{L}{2\pi}(E^\nu_{B}-E^0_{A})\ ,
\label{exp}
\end{equation}
For boundary potentials that do not lead to bound states one identifies the
exponents $x_0$ for the core-hole operator and $x_1$ for the core-hole
conduction-electron operator ($E^{0,1}_B$ being the ground state energies
in the $N$-($(N+1)$-)particle sector with B boundary conditions)
\cite{aflu:94}.  Fourier transforming (\ref{greenc}) the edge exponent in
(\ref{powerlaw}) is identified as
\begin{equation}
   \alpha=1-2\ x_1\ .
\label{edgeexp}
\end{equation}
In the presence of the various types of bound states the power-law
behaviour (\ref{powerlaw}) of $I(\omega)$ above the respective thresholds
can be determined by inserting the appropriate excited-state energy into
(\ref{exp}).  Finally, let us note that in the above discussion we have set
the Fermi velocities to one; the generalization to the two-component
Luttinger liquid with different Fermi velocities proceeds along the
same lines as in the case of periodic boundary conditions
\cite{izkr:89,frko:90}. 

In the remainder of the paper we follow the steps outlined above to
study the nature of the X-ray edge singularities in the $t$--$J$ and
Hubbard models for boundary terms $H_{1,L}$ chosen in such a way that
they preserve the integrability of these systems.

\section{The $t$--$J$ model}
\label{sec:tj}
In this section we determine the X-ray absorption exponents for a $t$-$J$
chain with the particular choice of core-hole potential described above. We
consider the following Hamiltonians \cite{essl:96}
\begin{eqnarray}
H&=& -{\cal P}\left(\sum_{j=1}^{L-1}\sum_\sigma c^\dagger_{j,\sigma}
c_{j+1,\sigma}+c^\dagger_{j+1,\sigma}c_{j,\sigma}\right){\cal P}
\nonumber\\
&&+2\sum_{j=1}^{L-1}{\vec S_j}\cdot{\vec S_{j+1}}
-\frac{n_jn_{j+1}}{4} +\sum_{j=1}^{L-1}n_j+n_{j+1} - \mu \hat{N}+
H_{\alpha\beta}\ ,
\label{hamil}
\end{eqnarray}
where ${\cal P}$ projects out double occupancies, $\vec{S_j}$ are spin
operators at site $j$,
$n_j=c^\dagger_{j,\uparrow}c_{j,\uparrow} + 
c^\dagger_{j,\downarrow}c_{j,\downarrow}$ and $n_j^h=1 -
n_{j,\uparrow} - n_{j,\downarrow}$. There are three different forms
for the boundary part $H_{\alpha\beta}$ of the Hamiltonian that are
compatible with integrability 
\begin{equation}
   H_{\tt aa} = h_1 n_1+h_Ln_L\ ,\qquad H_{\tt ba} = h_1 n_1
  + h_L (S^z_L-\frac{n_L^h}{2})\ ,\qquad H_{\tt bb} = h
  (S^z_1-\frac{n_1^h}{2}+S^z_L-\frac{n_L^h}{2})\ .  
\end{equation}
These correspond to localized potential ({\tt a}) and magnetic ({\tt b})
interactions of the conduction electrons with the disturbance due to the core
hole. Physically local magnetic field interactions are not very realistic; one
would rather expect a Kondo-like interaction which we cannot consider in the
present framework of integrable lattice models. In what follows we therefore
constrain our analysis to the model with {\tt aa} boundary conditions.  We
note that in the continuum limit $H_{\tt aa}$ gives rise to forward scattering
terms. We therefore expect that the X-ray exponents will generally not be
universal despite the fact that our boundary is perfectly reflecting in the
sense of Refs.~\onlinecite{prok:94,aflu:94}. As we will se below this is
indeed the case. However, the situation is somewhat more complicated than
this: unlike in the continuum limit \cite{prok:94,aflu:94} we do not impose
Neumann boundary conditions (on the lattice wave functions). The boundary
conditions should rather be thought of as being of mixed Dirichlet-Neumann
type ({\sl e.g.} $c\psi(0)+\partial_x\psi(0)=0$). The parameter $c$
enters the finite-size spectrum in the same way as the forward
scattering amplitude. Therefore in the continuum limit the forward
scattering amplitude is {\sl not} simply given by the boundary
chemical potential. As a result we recover the results of
Refs.~\onlinecite{prok:94,aflu:94} not for $h_{1,L}\to 0,1$ but for
some finite value that depends on the band filling (see below).

In the following we start by considering boundary fields in the
region $1\leq h\leq 2$. This is unphysical from the point of view of
the X-ray edge singularity where the potential due to the core hole
should be attractive but permits a pedagogical discussion of 
the formalism we use to calculate the finite-size energies necessary
for extracting X-ray exponents.

\subsection{Repulsive boundary fields: $1\leq h\leq 2$}
\label{sec:tjaa}

In this region of boundary fields holon boundary bound states
at both boundaries are present in the ground state of the $t$-$J$
chain. Defining  
\begin{equation}
  S_j = 2-\frac{2}{h_j}\ ,\quad j=1,L
\label{s}
\end{equation}
the Bethe Ansatz equations with respect to the reference state with
all spins up read \cite{essl:96}
\begin{eqnarray}
\left(e_1(\lambda_\alpha)\right)^{2L}&=&
  \prod_{\beta\neq\alpha}^{N_h+N_\downarrow}
  e_2(\lambda_\alpha-\lambda_\beta) e_2(\lambda_\alpha+\lambda_\beta) 
  \prod_{\gamma=1}^{N_h}
  e_{-1}(\lambda_\alpha-\lambda^{(1)}_\gamma) 
  e_{-1}(\lambda_\alpha+\lambda^{(1)}_\gamma)
\nonumber\\
1&=&e_{-S_1}(\lambda^{(1)}_\gamma)e_{-S_L}(\lambda^{(1)}_\gamma)
  \prod_{\beta=1}^{N_h+N_\downarrow}
  e_{1}(\lambda^{(1)}_\gamma-\lambda_\beta)
  e_{1}(\lambda^{(1)}_\gamma+\lambda_\beta)\ ,
\label{bae}
\end{eqnarray}
where $e_n(x)=\frac{x+\frac{in}{2}}{x-\frac{in}{2}}$.
The energy of a state corresponding to a solution of (\ref{bae}) is
\begin{equation}
E=h_1+h_L-\sum_{j=1}^{N_h+N_\downarrow}\frac{1}{\frac{1}{4}+\lambda_j^2}
+\mu N_h\ .
\label{bareE}
\end{equation}
We now observe that for $h_1>1$ solutions of (\ref{bae}) exists where (in the
thermodynamic limit) two roots $\lambda^{(1)}$ take the values
$-\frac{i}{2}S_1$ and $-\frac{i}{2}S_L$ respectively.  These roots correspond
to boundary bound states. The situation is analogous to the XXZ Heisenberg
chain studied in Ref.~\onlinecite{kask:96}. One finds that the ground state is
given by a distribution of roots such that both these boundary roots are
present. The logarithmic form of the Bethe equations (for a solution of
(\ref{bae}) with only real roots apart from the boundary roots) reads
\begin{eqnarray}
  \frac{2\pi}{L} I^s_\alpha &=& (2+\frac{1}{L}) \theta(\lambda_\alpha) -
  \frac{1}{L} \sum_{\beta}\theta(\frac{\lambda_\alpha-\lambda_\beta}{2})
  +\theta(\frac{\lambda_\alpha +\lambda_\beta}{2})
\nonumber\\
&&+ \frac{1}{L} \sum_{\gamma=1}^{N_h}
  \theta(\lambda_\alpha-\lambda^{(1)}_\gamma)
  +\theta(\lambda_\alpha+\lambda^{(1)}_\gamma)
  +\frac{\kappa(\lambda_\alpha)}{L}\
  ,\qquad \alpha=1\ldots N_\downarrow+N_h-1
\nonumber\\
  \frac{2\pi}{L} I^c_\gamma &=& \frac{1}{L}\sum_{\alpha}
  \theta(\lambda^{(1)}_\gamma-\lambda_\alpha)
  +\theta(\lambda^{(1)}_\gamma+\lambda_\alpha)
  + \frac{\omega_{}(\lambda^{(1)}_\gamma)}{L}\ ,\qquad
\gamma=1\ldots N_h-1
\label{baelog}
\end{eqnarray}
where $N_\downarrow$ is the number of electrons with spin down, $N_h$
is the number of holes, $I^{s,c}_\alpha$ are integer numbers,
$\theta(x)=2\arctan(2x)$ and 
\begin{eqnarray}
\kappa_{}(\l)&=&\theta(\frac{\l}{1+S_1})+\theta(\frac{\l}{1-S_1})
+\theta(\frac{\l}{1+S_L})+\theta(\frac{\l}{1-S_L})\ ,
\nonumber\\
\omega_{}(\l)&=&-\theta(\frac{\l}{S_1})-\theta(\frac{\l}{S_L})\ .
\label{bc}
\end{eqnarray}
In addition to (\ref{baelog}) we still have two equations determining the
precise values of the boundary roots. A detailed analysis of these
equations yields that the corrections to the thermodynamic values in a
finite system vanish exponentially with system size. This means that
for the purposes of the present work we can neglect these corrections.
We should note here that solutions of (\ref{baelog}) do not yield a {\sl
complete} set of states. For vanishing boundary fields such a basis
can be constructed by means of the $sl(1|2)$ symmetry of the
Hamiltonian \cite{essl:96}. For nonzero boundary fields this symmetry
is broken and we do not known how to complement the set of Bethe
states given by solutions of (\ref{baelog}). However, for the present
purposes this is not necessary: we are interested in the lowest energy
state in a particular sector of quantum numbers and it can be shown
that these states can always be obtained as solutions of
(\ref{baelog}) or the analogous equations based on the Bethe Ansatz
reference state with all spins down. We note that this ceases to be
true for the $t$-$J$ chain with {\tt ba} or {\tt bb} boundary terms.

The calculation of the finite-size spectrum proceeds along the lines of
Refs.~\onlinecite{essl:96} and \onlinecite{assu:97} so that we merely quote
the result
\begin{eqnarray}
   E^{(n)}&=& L e_\infty + f_\infty +\frac{\pi v_c}{L}\left\{
   \frac{1}{2}\frac{(\Delta N_c^0-\theta^c_{0})^2}{\xi^2}-\frac{1}{24}+N^c_+
   \right\}
\nonumber\\
   &&\qquad+\frac{\pi v_s}{L}\left\{
     (\Delta N_s^0-\frac{\Delta N_c^0}{2}-\theta^s_{0}
   +\frac{\theta^c_{0}}{2})^2-\frac{1}{24}+N^s_+\right\},
\label{fsop}
\end{eqnarray}
where $e_\infty$ is the ground state energy of the infinite system,
$v_c$ and $v_s$ are the Fermi velocities of holons and spinons
respectively, $\xi=\xi(\Lambda_c)$ is the dressed charge defined {\sl via}
\begin{eqnarray}
\xi(\lambda) &=& 1+\int_{-\Lambda_c}^{\Lambda_c} d\nu\ G_1(\lambda-\nu)\
\xi(\nu)\ , \nonumber\\
G_x(\lambda)&=&\frac{1}{2\pi}\int_{-\infty}^\infty d\omega\
e^{-i\omega\lambda} \frac{e^{-|x\frac{\omega}{2}|}}{2\cosh\frac{\omega}{2}}
=\frac{1}{2\pi}\Re e\left\lbrace\psi(\frac{3+x}{4}+i\frac{\lambda}{2})
-\psi(\frac{1+x}{4}+i\frac{\lambda}{2})\right\rbrace ,
\end{eqnarray}
where $\psi(x)$ is the digamma function. The integration boundary
$\Lambda_c$ is determined by the chemical potential (band filling). We
note that as we approach half-filling ($\mu\to 2\ln 2$)
$\Lambda_c\approx \sqrt{\frac{2}{3\zeta(3)}(2\ln 2-\mu)}$. 

\begin{figure}[ht]
\begin{center}
\noindent
\epsfxsize=0.55\textwidth
\epsfbox{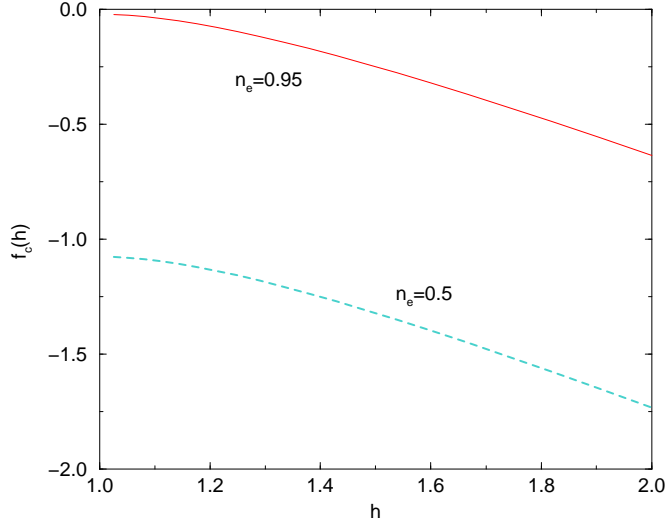}
\end{center}
\caption{\label{fig:fh}%
Energy of holon bound states as a function of boundary
chemical potential.}
\end{figure}

The term proportional to $N_+^\alpha=\sum_{\rm all\ pairs}
I_p^\alpha-I_h^\alpha$ is the contribution of particle-hole excitations, where
$I^\alpha_{p,h}$ are the integers corresponding to the roots of the particle
and the hole.  The quantities $\Delta N^0_c$ and $\Delta N^0_s$ denote the
deviations of the total particle number and the number of down spins from
their respective values for some reference state. This concept needs to be
introduced because in order to extract the X-ray exponents we need to compare
finite-size energies for {\sl different} boundary conditions. The state with
respect to which we measure the deviations of particle numbers is chosen such
that for the ground state $\Delta N^0_\alpha-\theta^\alpha_{0}=0$ for
$\alpha=c,s$. This may appear odd but turns out to be the most convenient
choice for calculating the energy difference between states with open and
closed boundary conditions. The quantities $\theta^{c,s}_0$ are
defined as

\begin{eqnarray}
\theta^\alpha_{0} &=& \frac{1}{2}\int_{-\Lambda_\alpha}^{\Lambda_\alpha}
d\nu\ \rho_\alpha^1(\nu)-\frac{1}{2}\ , \nonumber\\
\rho^1_\alpha(\lambda)&=&g_{\alpha,0}(\lambda) + 
\int_{-\Lambda_c}^{\Lambda_c}d\nu\left[\delta_{\alpha s}G_0(\lambda-\nu)
+\delta_{\alpha c}G_1(\lambda-\nu)\right]
\rho^1_c(\nu)\ ,\quad \alpha=c,s\ ,
\label{Adens}
\end{eqnarray}
where 
\begin{eqnarray}
g_{s,0}(\lambda)&=&\sum_{j=1,L}G_{S_j}(\lambda)+G_{|1-S_j|-1}+G_1(\lambda)\ ,
\nonumber\\
g_{c,0}(\lambda)&=&\sum_{j=1,L}G_{1+S_j}(\lambda)+G_{|1-S_j|}
-a_{S_j}(\lambda)-G_0(\lambda)\ . 
\end{eqnarray}
Last but not least the surface energy $f_\infty$ is found to be 
\begin{equation}
f_\infty= f_0 + f_c(h_1)+ f_c(h_L)\equiv f_0+f_{\rm bound}^{}\ ,
\end{equation}
where $f_0$ is the surface energy of the system in the absence of the
boundary bound states \cite{essl:96} and $f_c(h_j)$ are the
contributions of the holon boundary bound states. Note that these
contributions are of order one unless we fine-tune the boundary
fields. We find \cite{essl:96}
\begin{eqnarray}
f_0&=& - \frac{1}{2}\int_{-\Lambda_c}^{\Lambda_c} d\lambda\ 
   \varepsilon_c(\lambda)
   [a_{S_1}(\lambda)+a_{S_L}(\lambda)]-\frac{1}{2} 
   [\varepsilon_s(0)+\mu-2h_1-2h_L] \ ,
\nonumber\\
  f_c(h)&=&\mu-\pi(G_{3-\frac{2}{h}}(0)+G_{-1+\frac{2}{h}}(0))+\frac{1}{2}
  \int_{-\Lambda_c}^{\Lambda_c}d\nu\ [G_{3-\frac{2}{h}}(\nu) +
  G_{-1+\frac{2}{h}}(\nu)]\ \varepsilon_c(\nu)\ ,
\label{E}
\end{eqnarray}
where $a_x(\lambda) = \frac{1}{2\pi} \frac{x}{\lambda^2+\frac{x^2}{4}}$ and
where the dressed energies are given as solutions of
\begin{eqnarray}
  \varepsilon_s(\lambda)&=&-2\pi
G_0(\lambda)+\int_{-\Lambda_c}^{\Lambda_c}d\nu\ G_0(\lambda-\nu)\
\varepsilon_c(\nu)\ , \nonumber\\
  \varepsilon_c(\lambda)&=&\mu-2\pi
G_1(\lambda)+\int_{-\Lambda_c}^{\Lambda_c}d\nu\ G_1(\lambda-\nu)\
\varepsilon_c(\nu)\ . 
\label{dresseden}
\end{eqnarray}
The bound state energy $f_c(h)$ as a function of boundary chemical
potential is shown for different band fillings in Fig.~\ref{fig:fh}.
This characterizes the relevant part of the low-lying finite-size
spectrum of the open $t$-$J$ chain with boundary fields in the sector
where $N_\uparrow\geq N_\downarrow$. In order to extract the X-ray
exponents we need to consider states with $N_\downarrow\geq
N_\uparrow$ for the case where the core electron carries spin
down. This can be taken care of by changing the reference state of the
Bethe Ansatz to the state with all spins down \cite{essl:96}. The
result is of the same form as (\ref{fsop}) but with redefined $\Delta
N_s^0$.  

We also need the finite-size ground state energy of the $t$-$J$ model
with periodic boundary conditions. It is given by \cite{kaya:91}
\begin{equation}
   E^{(0)}= L e_\infty -\frac{\pi}{6L}(v_c+v_s)\ .
\label{fsper}
\end{equation}

We now have the necessary machinery to determine X-ray exponents. One
should keep in mind that we presently have {\sl repulsive} boundary
fields. For pedagogical reason we nonetheless will calculate X-ray
exponents for this case: 

\vskip.3cm
\underbar{\sc Absolute Threshold:}\indent
The lowest (in frequency) threshold in the X-ray absorption
intensity occurs at some frequency $\omega_0$ and is associated with
an intermediate state in which both holon bound states are
occupied. For the case where the core electron has spin up this
corresponds to $\Delta N^0_c=-3\ ,\ \Delta N^0_s=-1$. Combining
(\ref{fsop}), (\ref{fsper}), (\ref{exp}) and (\ref{edgeexp}) we obtain
\begin{equation}
   \alpha_{abs} = \frac{1}{2}-\frac{[3+\theta_{0}^c]^2}{2\xi^2}\ .
   \label{x0}
\end{equation}
For the case where the core electron has spin down we need to proceed
as outlined above and use the Bethe Ansatz solution with a different
reference state. The final result is the same as (\ref{x0}) as $H_{\tt
aa}$ preserves the discrete spin reversal symmetry.

\vskip.3cm
\underbar{\sc Intermediate Thresholds:}\indent
The second and third thresholds occur when one of the holon bound
states is occupied but the other one is not. Let us consider the case
where the bound state at $1$ is occupied. 
The corresponding threshold in the X-ray absorption rate is at
$\omega_0-f_c(h_1)$. As only one holon bound state is
occupied we now have $\Delta N^0_s=-1,\ \Delta N^0_c=-2$ and the expressions for
the quantities $\theta^{s,c}$ in (\ref{fsop}) get modified. They are now
given by (\ref{Adens}) but with different driving terms
\begin{equation}
\theta_{1}^s=\frac{\theta_{1}^c}{2} + \frac{1}{2}\ ,\quad
g_{c,1}(\lambda)=G_{1+S_1}(\lambda)+G_{1-S_1}(\lambda)
-G_0(\lambda)-a_{S_1}-a_{s_L}\ . 
\end{equation}
The X-ray exponent associated with this threshold is
\begin{equation}
\alpha_{int} = \frac{1}{2}-\frac{[2+\theta_{1}^c]^2}{2\xi^2} \ .
\end{equation}
\vskip.3cm
\underbar{\sc Band Threshold:}\indent
The fourth and final threshold occurs at $\omega_0-f_c(h_L)-f_c(h_1)$
when neither bound state is occupied. This corresponds to the case
where the core electron is emitted into the conduction band where it
decomposes into an antiholon and a spinon. Then $\Delta N^0_c=-1\ ,\
\Delta N^0_s=-1$, $\theta^c_{3}=2\theta^s_{3}$,
$g_{c,3}(\lambda)=-G_0(\lambda)-a_{S_1}(\lambda) -a_{S_L}(\lambda)$ and the associated
X-ray exponent is 
\begin{equation}
\alpha_{band} = \frac{1}{2}-\frac{(1+\theta_{3}^c)^2}{4\xi^2}\ .
\label{x3}
\end{equation}

\subsection{Attractive Boundary fields: $0\leq h\leq 1$}
\begin{figure}[ht]
\noindent
\epsfxsize=0.45\textwidth
(a)
\epsfbox{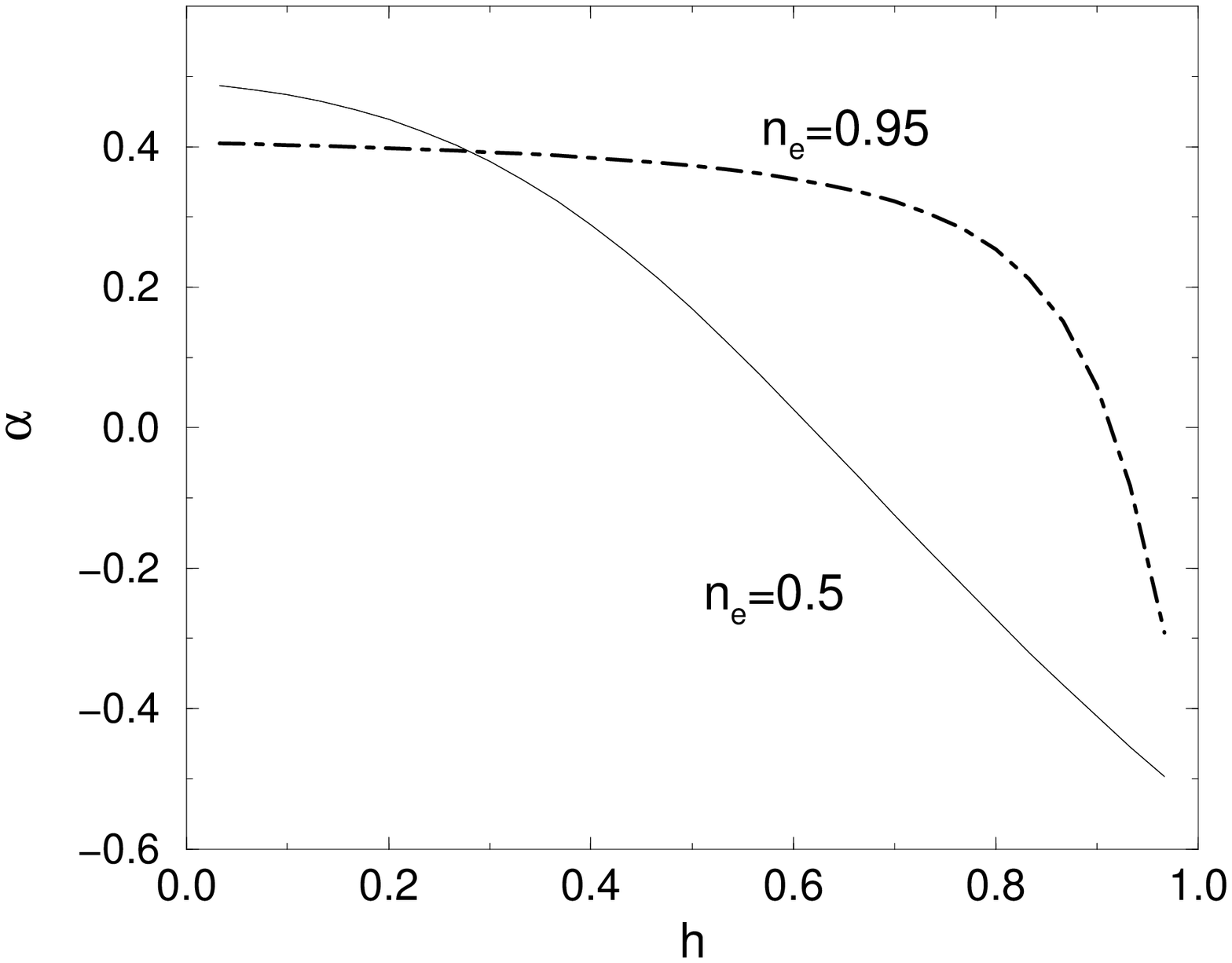}
\epsfxsize=0.45\textwidth
\hfill
(b)
\epsfbox{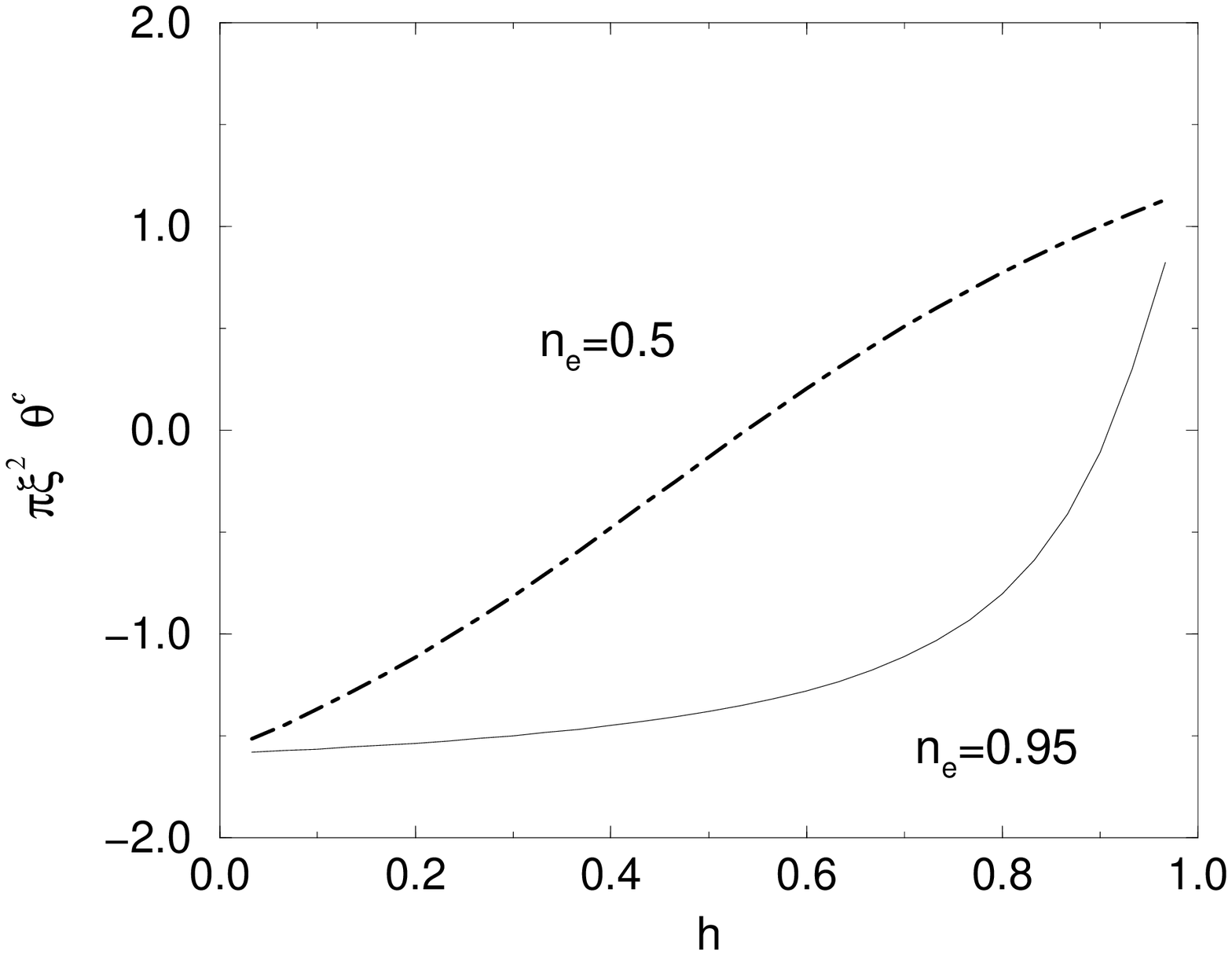}
\caption{\label{fig:weak}%
X-ray exponents (a) and ${\pi\theta^c}/{\xi^2}$ (b) for the $t$-$J$
model with {\tt aa} boundary conditions and $h_1=h_L=h<1$.}
\end{figure}

This region of boundary fields corresponds to an attractive core hole
potential because of the form of the third last term in (\ref{hamil}).
Now no boundary bound states exist. The analysis of the finite-size
spectrum follows the one above, the only difference being the absence
of purely imaginary roots.  The X-ray exponent is of the same form as
(\ref{x3}) where we should keep in mind that $S_{1,L}$ now are
negative. The results for two different band fillings are plotted in
Fig.~\ref{fig:weak}a as functions of the boundary chemical potential
$h=h_1=h_L$.
Our result coincides with Refs.~\onlinecite{prok:94,aflu:94} if we make the
identification $\theta^c_{}=\frac{V_f}{\pi}\xi^2$, where $V_f$ is the forward
scattering amplitude of the core hole potential in the continuum limit. We see
that $\theta^c$ does not vanish for $h_{1,L}\to 0$. As explained above the
continuum $V_f$ is not simply given by the boundary chemical potential so that
there is no contradiction. In Fig.~\ref{fig:weak}b we plot
${\pi\theta^c}/{\xi^2}$ as a function of $h$.

\subsection{Attractive Boundary fields: $h\leq 0$}

In this range of boundary chemical potential the analysis of the
finite-size spectrum is less intuitive than above. The Bethe equations
(\ref{bae}) allow a variety of boundary string solutions like the ones
encountered in the repulsive case. However one finds that none of
these complex roots is present in the ground state. We interpret this
as follows: in the ground state antiholons and spinons are bound to
the boundaries. States where some of these bound states are unoccupied
are characterized by imaginary roots of the Bethe equations. In
support of this interpretation we can compute the particle number at
the boundary site. It is given by $\frac{\partial E}{\partial
h}$, where $h$ is the boundary field. We find that in the ground state
there is a strong enhancement of charge at the boundary site as
compared to the bulk. The states involving imaginary roots of the
Bethe equations exhibit a significant decrease in charge at the
boundary as compared to the ground state, which is consistent with our
interpretation. 

\vskip.3cm
\underbar{\sc Absolute Threshold:}\indent

In order to calculate the X-ray exponent for the lowest threshold we
need the finite-size energy of the ground state for $h<0$. As no
complex roots of the Bethe equations are present the analysis is
straightforward and very similar to the band threshold for $2>h>1$. We
find
\begin{equation}
\alpha_{abs}=\frac{1}{2}-\frac{(1+\theta^c)^2}{2\xi^2}\ ,
\end{equation}
where $\theta^c$ is given by (\ref{Adens}) with $g_c(\lambda)=
-G_0(\lambda)-a_{S_1}(\lambda)-a_{S_L}(\lambda)$.

\begin{figure}[ht]
\begin{center}
\noindent
\epsfxsize=0.6\textwidth
\epsfbox{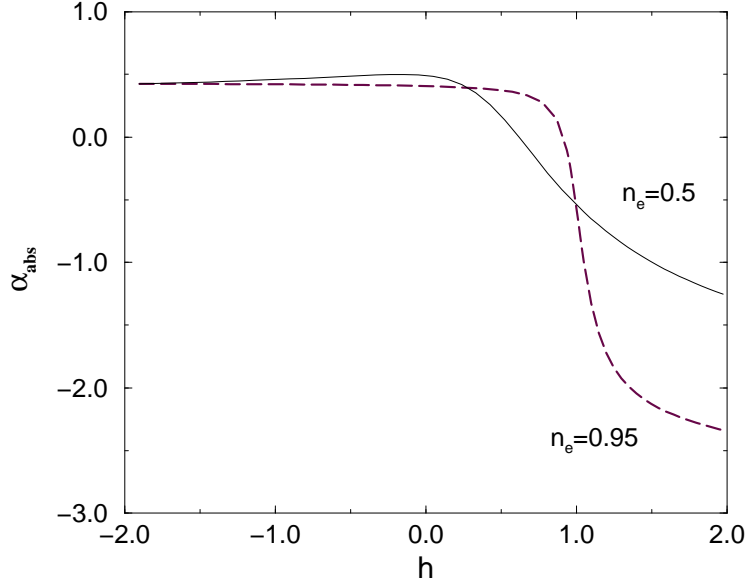}
\end{center}
\caption{\label{fig:alphaabs}%
X-ray exponents for the absolute threshold in the $t$-$J$ model with
$h_1=h_L=h$ at almost half-filling and quarter-filling.}
\end{figure}

In Fig.~\ref{fig:alphaabs} the X-ray exponents of the absolute
threshold are plotted as functions of the boundary chemical potential
for two different band fillings. For simplicity we only consider the
case $h_1=h_L=h$. 
We see that in the physical regime $h\leq 1$ there is always a
singularity associated with the absolute threshold {\sl i.e.}
$I(\omega)$ always diverges. 

\vskip.3cm
\underbar{\sc Higher Thresholds:}\indent
Let us consider the case in which two complex roots $\lambda^{(1)}$
are present and take the values $-\frac{i}{2}S_{1,L}$ respectively.
The Bethe equations read
\begin{eqnarray}
   &&\left(e_1(\lambda_\alpha)\right)^{2L}\prod_{j=1,L}e_{1-S_j}(\lambda_\alpha)e_{1+S_j}(\lambda_\alpha)
\nonumber \\
   &&\qquad=\prod_{\beta\neq\alpha}^{N_h+N_\downarrow}
   e_2(\lambda_\alpha-\lambda_\beta) e_2(\lambda_\alpha+\lambda_\beta) 
   \prod_{\gamma=1}^{N_h-2}
   e_{-1}(\lambda_\alpha-\lambda^{(1)}_\gamma) 
   e_{-1}(\lambda_\alpha+\lambda^{(1)}_\gamma)
\nonumber\\
   &&1=e_{-S_1}(\lambda^{(1)}_\gamma)
   e_{-S_L}(\lambda^{(1)}_\gamma)\prod_{\beta=1}^{N_h+N_\downarrow}
   e_{1}(\lambda^{(1)}_\gamma-\lambda_\beta)
   e_{1}(\lambda^{(1)}_\gamma+\lambda_\beta)\ .
\label{baeattr}
\end{eqnarray}
Following through the same steps as before we find that this state has
a gap of magnitude $\Delta f = f_c(S_1) + f_c(S_L)$ where
\begin{equation}
f_c(S)=\mu-\pi[G_{S+1}(0)-G_{S-1}(0)]+\frac{1}{2}\int_{-\Lambda_c}^{\Lambda_c}
d\lambda\ \varepsilon_c(\lambda)[G_{S+1}(\lambda)-G_{S-1}(\lambda)]\ .
\end{equation}
We interpret this state as differing from the ground state by leaving
boundary bound states of antiholons unoccupied. Consequently we find a
threshold in the X-ray absorption probability at a frequency $\Delta
f$ higher than the absolute threshold with exponent
\begin{equation}
\alpha_{int}=\frac{1}{2}-\frac{(3+\theta^c)^2}{2\xi^2}\ ,
\end{equation}
where $\theta^c$ is given by (\ref{Adens}) with $g_c(\lambda)=
-G_0(\lambda)+\sum_{j=1,L}G_{S_j+1}(\lambda)-G_{S_j-1}(\lambda)
-a_{S_j}(\lambda)$.

Thresholds at lower frequencies occur if we have only one imaginary
root $\lambda^{(1)}=-\frac{i}{2}S$ where $S$ is either $S_1$ or
$S_L$. The corresponding states have a gap equal to $\Delta f= f_c(S)$
and give rise to exponents
\begin{equation}
\alpha_{int}^\prime=\frac{3}{4}-\frac{(2+\theta^c)^2}{2\xi^2}\ ,
\end{equation}
where $\theta^c$ is given by (\ref{Adens}) with $g_c(\lambda)=
-G_0(\lambda)+G_{S+1(\lambda)}-G_{S-1}(\lambda)-a_{S_1}(\lambda)
-a_{S_L}(\lambda)$. A numerical
solution of the relevant integral equations for a quarter filled band
shows that $\alpha_{int}$ is negative and therefore leads to a
``shoulder'' in $I(\omega)$ as in Fig.~\ref{fig:int} b). On the other
hand we find that $\alpha^\prime_{int}$ is positive and leads to a
singularity. 

The cases imvestigated above by no means exhaust the list of states
with imaginary roots. For 
example there is a state with two imaginary $\lambda^{(1)}$'s taking
the values $-\frac{i}{2}S_{1,L}$ and two imaginary $\lambda$'s taking
the values $\frac{i}{2}(1-S_{1,L})$ respectively. This type of
solution of the Bethe equation also gives rise to three thresholds as
imaginary $\lambda$'s are only allowed if their respective ``partner''
$\lambda^{(1)}$ is present as well. The calculation of the X-ray
exponents is completely analogous to the case treated above so that we
omit it.

\section{The Hubbard model}
The one dimensional Hubbard model with open boundary conditions of type
{\tt aa} (i.e.\ boundary chemical potentials only)
\begin{equation}
  H   = - \sum_{j=1}^{L-1} \sum_\sigma\left(c^\dagger_{j,\sigma}c_{j+1,\sigma}
        +c^\dagger_{j+1,\sigma}c_{j,\sigma}\right)
        + 4u\sum_{j=1}^L n_{j\uparrow}n_{j\downarrow}
        + \mu \hat{N} - h_1 n_1 - h_L n_L\ 
\label{ham:hubb}
\end{equation}
is soluble by means of the Bethe Ansatz as shown in
Refs.~\onlinecite{hschulz:85,assu:96} (note that the boundary potentials are
defined in a different way than above: to identify $h_{1,L}$ in (\ref{ham:hubb}) with
those used for the $t$--$J$ model one should replace $h_{1,L}\to1-h_{1,L}$).
Applying boundary magnetic fields instead also leaves the Hubbard model
integrable \cite{BCHubb} but will not be considered here.  The Bethe
Ansatz equations determining the spectrum of (\ref{ham:hubb}) in the
$N_{e}$-particle sector with magnetization $M={1\over2}N_{e} -N_{\downarrow}$
read \cite{hschulz:85,assu:96}
\begin{eqnarray}
  &&e^{2ik_{j}L} B^{(1)}_c(k_j) B^{(L)}_c(k_j) =
   \prod_{\beta=1}^{N_\downarrow} e_{2u}(\sin k_{j}-\lambda_{\beta})
      e_{2u}(\sin k_{j}+\lambda_{\beta})\ , \qquad j=1,\ldots,N_{e}
\nonumber \\
  &&B^{(1)}_s(\lambda_\alpha) B^{(L)}_s(\lambda_\alpha)
  \prod_{j=1}^{N_e} e_{2u}(\lambda_{\alpha}- \sin k_{j})
                    e_{2u}(\lambda_{\alpha}+ \sin k_{j})
\nonumber \\
  &&\qquad =\prod_{\beta\ne\alpha}^{N_{\downarrow}}
                    e_{4u}(\lambda_{\alpha}-\lambda_{\beta})
                    e_{4u}(\lambda_{\alpha}+\lambda_{\beta})\ ,
                        \qquad \alpha=1,\ldots,N_{\downarrow}.
\label{bae:h}
\end{eqnarray}
The quasi momenta $k_j$ and the spin rapidities $\lambda_\alpha$ paramatrize
an eigenstate of (\ref{ham:hubb}) with energy
\begin{equation}
  E = \mu N_{e}-2\sum_{j=1}^{N_e}\cos k_j\ .
  \label{energy:hubb}
\end{equation}
For small values of the boundary fields the ground state configuration is
given by distributions of {\em real} $k_j$ and $\lambda_\alpha$ and
\begin{equation}
   B^{(x)}_c(k) = \left({{e^{ik}-h_x}\over{1-h_x e^{ik}}}\right)\ ,
   \qquad
   B^{(x)}_s(\lambda) = 1\ ,
\label{bou0:h}
\end{equation}
contain the phase shifts due to the boundaries (this case has been discussed
in Ref.~\onlinecite{assu:96}).  For sufficiently large boundary chemical
potentials $h_{1,L}$, however, the Bethe Ansatz equations (\ref{bae:h}) allow
for various complex solutions corresponding to boundary bound states for
antiholons, spinons and pairs of electrons, respectively \cite{befr:pp}:
First, for $h_{1,L}>1$ one finds bound states parametrized by $k=i\ln h_{1,L}$
with exponential accuracy in the thermodynamic limit $L\to\infty$.  The quasi
momenta parametrize the charge part of the states: hence this solution
corresponds to a charge (or antiholon) bound to the surface.  Inserting this
solution in the second set of Eqs.~(\ref{bae:h}) leads to a boundary phase
shift in addition to the product over the real quasi momenta $k_j$ which
modifies $B_s$ ($B_c$ remains unchanged):
\begin{equation}
  B^{(x)}_s(\lambda) = e_{2(u+S_x)}(\lambda)\ e_{2(u-S_x)}(\lambda)\ ,
\label{bou1:h}
\end{equation}
where we have introduced $S_x=(h_x-1/h_x)/2>0$ with $x=1$ or $L$.
Analyzing the resulting equations we find that a new type of solution
arises at $S_x=u$, i.e.\ $h_x=u+\sqrt{u^2+1}$: Beyond this point a complex
solution $\lambda=i(S_x-u)$ for the spin rapidities is allowed. We
note that spinons are to be identified with holes in the distribution
of spin rapidities. Again, occupation of this state modifies the
boundary phase shifts $B_{c,s}$:
\begin{eqnarray}
  B^{(x)}_c(k) &=& \left({{e^{ik}-h_x}\over{1-h_x e^{ik}}}\right)
        e_{-2S_x}(\sin{}k_j) e_{2(S_x-2u)}(\sin{}k_j)\ ,
\nonumber \\
  B^{(x)}_s(\lambda) &=& e_{2(S_x-3u)}(\lambda)\ e_{2(u-S_x)}(\lambda)\ .
\label{bou2:h}
\end{eqnarray}
Finally, boundary potentials with $S_x>2u$ can bind a (singlet) pair of
electrons to site $x$.  Such a state is parametrized by two complex quasi
momenta $\sin{}k_0^{(\pm)}=\lambda_0\pm iu$ and a single complex spin
rapidity $\lambda_0=i(S_x-u)$ as before.  The remaining real solutions of
the Bethe Ansatz equations are determined by (\ref{bae:h}) with
\begin{equation}
  B^{(x)}_c(k) = \left({{e^{ik}-h_x}\over{1-h_x e^{ik}}}\right)
        e_{-2S_x}(\sin{}k_j) e_{2(S_x-2u)}(\sin{}k_j)\ ,
\qquad
  B^{(x)}_s(\lambda)=1\ .
\label{boup:h}
\end{equation}

Depending on the strength of the boundary potential we have to
distinguish between the following cases in order to describe
the spectrum: in addition to the case discussed in
Ref.~\onlinecite{assu:96}, where the solution of the Bethe Ansatz eqs.\ is
given in terms of real $k_j$ and $\lambda_\alpha$ only, one can find either
\begin{itemize}
\item
an antiholon in a bound state (corresponding to a complex $k$) and the
spinon in the corresponding band (which implies the presence of a complex
$\lambda$ for $S_x>u$),

\item
an antiholon {\em and} a spinon in bound states (parametrized by a complex
$k$ for $S_x>u$),

\item
and finally, for $S_x>2u$, a pair of electrons bound by the potential.
\end{itemize}
Each of these configurations gives rise to a continuous spectrum above
a threshold that depend on the occupation of the boundary states.

In the following, we shall discuss some of these cases for the symmetric
choice $h_1=h_L=h$ of the boundary potentials.  The bound states discussed
above will occur pairwise at the given thresholds (corresponding to sites
$1$ and $L$, respectively).  As for the $t$--$J$ model we shall consider
the logarithmic form of the Bethe Ansatz equations (\ref{bae:h}) for low
lying states above these thresholds:
\begin{eqnarray}
  {2\pi I_j \over L} &=& 2k_j + {1\over L}\sum_{\beta=1}^{M}
        \left\{
        \theta\left({\sin k_j-\lambda_\beta\over 2u}\right) +
        \theta\left({\sin k_j+\lambda_\beta\over 2u}\right) \right\}
        + {1\over L}\kappa(k_j)\ ,
        \qquad j=1,\ldots,N
\nonumber \\
  {2\pi J_\alpha \over L} &=& 
        {1\over L} \sum_{j=1}^{N} 
        \left\{
        \theta\left({\lambda_\alpha-\sin k_j\over 2u}\right) +
        \theta\left({\lambda_\alpha+\sin k_j\over 2u}\right) \right\} 
\nonumber \\
        &&- {1\over L} \sum_{\beta\ne\alpha}^{M}
        \left\{
        \theta\left({\lambda_\alpha-\lambda_\beta\over 4u}\right) +
        \theta\left({\lambda_\alpha+\lambda_\beta\over 4u}\right) \right\}
        + {1\over L}\omega\left(\lambda_\alpha\right)\ ,
        \qquad \alpha=1,\ldots,M\ .
\label{baelog:h}
\end{eqnarray}
Here the summations extend over the real roots $k_j$ and $\lambda_\alpha$.
The functions $\kappa$ and $\omega$ contain the phase shifts due to the
boundary fields and occupation of the boundary bound states.

\subsection{Band threshold}
\label{sec:hband}
\begin{figure}[ht]
\begin{center}
\noindent
\epsfxsize=0.6\textwidth
\epsfbox{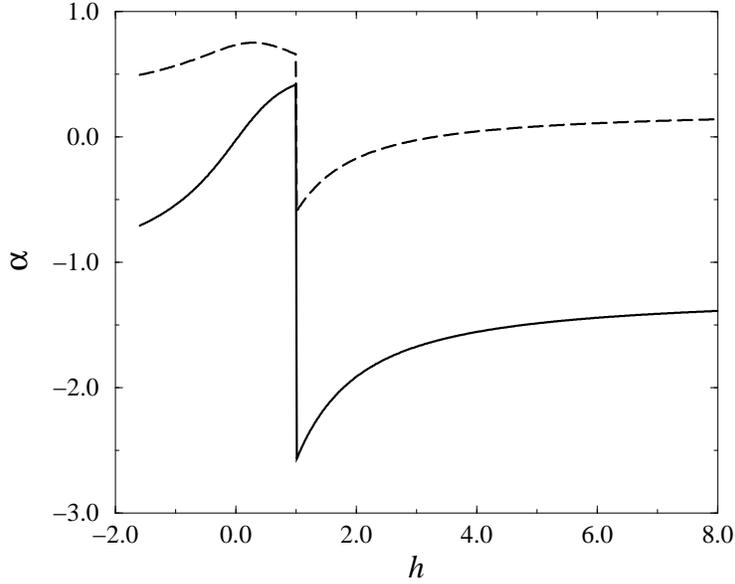}
\end{center}
\caption{\label{fig:hubb_b}%
X-ray edge exponents for band absorption (full line) and photoemission
(dashed line) in the Hubbard model as a function of the boundary chemical
potential $h$ for $u=1$, $n_e=0.5$.
}
\end{figure}

The edge singularity with the highest threshold corresponds to excitation
in states with no bound states occupied by the particles.  This situation
was studied in Ref.~\onlinecite{assu:96}. Like in the case of the $t$--$J$
model this does in fact imply the occupation of a holon bound state for
repulsive boundary potentials $h_{1,L}<-1$: computation of the particle
number on the boundary site shows a depletion due to the presence of the
holon\cite{befr:pp}.
In the Bethe Ansatz equations the only boundary phase shifts are those due
to the boundary potentials, i.e.\ (\ref{bou0:h}).  The resulting
$\kappa(k)$ in (\ref{baelog:h}) is given by
\begin{equation}
  \kappa(k) = -2i \ln\left({{e^{ik}-h}\over{1-he^{ik}}}\right) ,
\end{equation}
while $\omega(\lambda)=0$.  The finite size spectrum for the relevant
boundary conditions is again given by (\ref{fsop}) and (\ref{fsper})
\cite{woyn:89,assu:96}.  The dressed charge $\xi=\xi(Q)$ for the Hubbard
model is defined in terms of the solution of the integral equation ($Q$
varies between $0$ and $\pi$ as a function of the density of electrons and
the coupling constant) \cite{woyn:89,frko:90}
\begin{eqnarray}
  \xi(k) &=& 1+\int_{-Q}^{Q} dk' \cos{}k'\ 
        {\bar K}(\sin{}k-\sin{}k') \xi(k')\ ,
\nonumber\\
  {\bar K}(x) &=& {1\over2\pi}\int_0^\infty d\omega 
        {e^{-u\omega}\over\cosh{}u\omega}\cos{}\omega x\ .
\label{dressC:h}
\end{eqnarray}
Here $\theta^{c,s}$ are related {\sl via} $\theta^s={1\over2}\theta^c$ with
\begin{equation}
  \theta^c = {1\over2} \left(\int_{-Q}^Q dk \rho_c^{(1)}(k)\ -1\right)
\label{thc:h}
\end{equation}
for our choice of the reference state.  The $O(1/L)$ contribution
$\rho_c^{(1)}$ to the density from the boundary fields is given in terms of
the integral equation
\begin{equation}
  \rho_c^{(1)}(k) = {\bar g}_c(k) + \cos{}k\ \int_{-Q}^{Q} dk' 
           {\bar K}(\sin{}k-\sin{}k') \rho_c^{(1)}(k')\ .
\label{irho:h}
\end{equation}
For the case considered here the driving term in this equation is found to
be (after integrating out the spinon-part of the densities)
\begin{equation}
 {\bar g}_c(k) = {\bar g}_c^{(0)}(k) =
         {1\over\pi}\ {1-h^2\over 1+h^2-2h\cos{}k}
        -{\cos{}k \over 4u\cosh({\pi\over2u}\sin{}k)}\ .
\label{drv0:h}
\end{equation}
An analytic solution of this integral equation is possible in certain
limits only.  It simplifies essentially in the strong coupling limit where
${\bar K}(x)\equiv \ln2/2\pi u$.  This allows to give a simpler expression
for $\theta^c$ in terms of the driving term
\begin{equation}
  \theta^c \simeq {1\over 2} \left[ \left(1+{\ln2\over\pi u}\sin Q\right)
                  \int_{-Q}^Q dk {\bar g}_c(k)\ -1 \right]\ 
  \quad{\mathrm for~} u\to\infty\ .
\label{thinf:h}
\end{equation}
Furthermore it is known that $Q=\pi n_c$ and $\xi=1$ in this limit.  With
(\ref{drv0:h}) we find
\begin{equation}
   \theta^c = {2\over\pi}\arctan\left({1+h\over1-h}\tan{\pi
   n_c\over2}\right) -{1\over2}\
\label{thinf0}
\end{equation}
for infinite coupling \cite{frzv:97}.
In general the integral equations have to be solved numerically to compute
the X-ray edge exponents from (\ref{exp}) by comparing (\ref{fsop}) to the
finite size ground state energy of the Hubbard chain with periodic boundary
conditions (\ref{fsper}).  For absorption of the core electron into the
band we have to choose $\Delta N_c^0=1$.  The number of down spins in the
system changes by $\Delta N_s^0=0$ or $1$ depending on the spin of the core
electron.  Without magnetic fields the Bethe Ansatz states are highest
weight in the spin $SU(2)$, i.e.\ correspond to the first case.  This
results in the following expression for the exponent
\begin{equation}
  \alpha_{band} = {1\over2} - {1\over 2\xi^2} \left(\theta^c-1\right)^2\ .
\label{expo_band:h}
\end{equation}
{}From (\ref{thinf0}) we find that there is a discontinuity of
$\alpha_{band}(h)$ at $h=1$: at this point the charge bound state first
appears leading to a jump of the exponent from ${3/8}$ to ${-21/8}$ at
$u=\infty$ (note that small negative exponents correspond to a `shoulder'
rather than a singularity in the absorption profile \cite{cono}, exponents
$\alpha<-1$ will hardly lead to an observable feature).  Large boundary
potentials $h\to\pm\infty$ lead to $\theta^c\to-(n_c+{1\over2})$ in the
strong coupling limit giving $\alpha_{band}\to -{1\over2} (n_c^2+3n_c+
{5\over4})$ which is {\em always} negative.  Numerical solutions of the
equations show a similar behaviour for finite $u$ (see
Fig.~\ref{fig:hubb_b}).

Similarly, the singularity of the absorption intensity measured in a 
photoemission experiment is given by a power law with exponent obtained from
(\ref{exp}) with $\Delta N_{cs}^0=0$:
\begin{equation}
  \alpha_{photo} = {3\over 4} - {1\over2\xi^2}
\left(\theta^c\right)^2\ ,
\label{expo_photo:h}
\end{equation}
which exhibits a jump from $5/8$ to $-3/8$ at $h=1$ and approaches
${5\over8}-{1\over2}n_c(n_c+1)$ at $h\to\infty$ for infinite coupling.
Note that $(1/2\xi^2)$ varies as a function of the bulk density $n_e$ of
electrons and the interaction strength between $1/4$ for noninteracting
fermions and $1/2$ in the infinite $u$ limit of the Hubbard model
\cite{frko:90}, while $\theta^c$ contains the dependence on the strength of
the boundary potentials $h_{1,L}$ (in addition to $n_c$ and $u$).

For weak boundary fields $h<1$ these expressions coincide with those found
in the framework of a bosonized theory of spin carrying electrons
\cite{aflu:94,prok:94} provided that we identify $\theta^c$ with the
forward scattering amplitude of the core hole potential (see also the
discussion at the beginning of Sect.~\ref{sec:tj}).

\subsection{Absolute threshold}
\label{sec:habs}
Let us now consider X-ray processes which excite the system into the sector
with {\em all} bound states occupied, i.e.\ the absolute threshold for
absorption.  Following the discussion above one has to distinguish four
cases: For sufficiently small boundary fields ($h<1$) there are no bound
states, which is the situation considered in the previous section.

For boundary fields $1<h<u+\sqrt{u^2+1}$ a charge can be bound to either
boundary.  This changes the boundary phase shifts according to
(\ref{bou1:h}).  The computation of the finite size spectrum is complete
analogeous to the case considered above and results in (\ref{fsop}).  The
shifts of the numbers $\Delta N_{cs}^0$ are now found to be $\theta^{s}=
{1\over2}\theta^c+1$ and $\theta^c$ again given by (\ref{thc:h}).  The
different boundary phase shifts modify the driving term in (\ref{irho:h})
to
\begin{equation}
  {\bar g}_c(k) = {\bar g}_c^{(0)}(k) + \cos{}k\ f_b(\sin{}k)
\label{gc1:h}
\end{equation}
with
\begin{equation}
   f_b(x) = 2 a_{2(2u-S)}(x) + 
        {1\over u}\left\{ G_{1+{S\over u}}\left({x\over2u}\right) 
                     - G_{3-{S\over u}}\left({x\over2u}\right)
             \right\}\ .
\label{fb1:h}
\end{equation}
For the computation of the edge exponent from (\ref{fsop}) we have to
choose $\Delta N_c^0=-1$ (the number of charges in the band is increased by
one due to the absorption of the core electron, but at the same time two of
the band electrons occupy the bound states in the final state).  With
$\Delta N_s^0=0$ as before one obtains
\begin{equation}
  \alpha_{abs} = {1\over2} - {1\over 2\xi^2}\left(\theta^c+1\right)^2\ .
\label{expo_abs1:h}
\end{equation}

Increasing the boundary potentials such that $u+\sqrt{u^2+1} < h <
2u+\sqrt{4u^2+1}$ the Bethe Ansatz state of lowest energy is contains both
complex $k$ {\em and} complex $\lambda$ leading to phase shifts
(\ref{bou2:h}).  As discussed above this corresponds to occupied charge
bound states while the spinon bound states are empty.  Analysing the Bethe
Ansatz equations we find $\theta^s={1\over2}\theta^c-1$.  The function
$\rho_c^{(1)}(k)$ is determined by the same set of equations
(\ref{irho:h}), (\ref{gc1:h}) and (\ref{fb1:h}) as above.  The state
relevant for the edge exponent is now determined by the quantum numbers
$\Delta N_c^0=-1$ and $\Delta N_s^0=-2$ which gives again
(\ref{expo_abs1:h}).

A final change in the configuration describing the absolute ground
state occurs for $h>2u+\sqrt{4u^2+1}$ ($S>2u$).  The presence of bound
pairs of electrons leads to the phase shifts (\ref{boup:h}) in
the Bethe Ansatz equations. The quantities determining the edge exponents
are now $\theta^s={1\over2}\theta^c$, where $\theta^c$ has to be computed
from (\ref{irho:h}) with
\begin{equation}
  {\bar g}_c(k) = {\bar g}_c^{(0)}(k) + 2\cos{}k 
     \left\{ a_{2S}(\sin{}k) - a_{2(S-2u)}(\sin{}k) \right\}\ .
\end{equation}
The quantum numbers of the final state are $\Delta N_c^0=-3$ and $\Delta
N_s^0=-2$ which gives
\begin{equation}
  \alpha_{abs} = {1\over2} - {1\over2\xi^2}\left(\theta^c+3\right)^2\ .
\label{expo_ab2p:h}
\end{equation}

Again, the equations simplify significantly in the strong coupling limit
where one should rescale $S$ by $u$ to see the different regimes.  Using
(\ref{thinf:h}) we can combine Eqs.\ (\ref{expo_band:h}),
(\ref{expo_abs1:h}) and (\ref{expo_ab2p:h}) into
$\alpha_{abs}={1\over2}\left(1-x^2\right)$ where
\begin{equation}
  x = {2\over\pi}\left\{
        \arctan\left({{h-\cos\pi n_c}\over{\sin\pi n_c}}\right)
       +\arctan\left({{h-4u}\over{2\sin\pi n_c}}\right)
    \right\}-n_c+{1\over2}\ .
\end{equation}
Hence we find the following expression for the edge exponent of the
absolute threshold in the strong coupling limit
\begin{equation}
   \alpha_{abs} \to 
      \left\{ \begin{array}{cl}
        -{1\over8} (2n_c+5)(2n_c+1)  & {\mathrm~for~} h\ll-1 \\[4pt]
        -{1\over8} (2n_c+1)(2n_c-3)  & {\mathrm~for~} 1\ll h\ll4u \\[4pt]
        -{1\over8} (2n_c-3)(2n_c-7) & {\mathrm~for~} h\gg4u
                            \end{array} \right.\ .
\end{equation}

\begin{figure}[t]
\begin{center}
\noindent
\epsfxsize=0.6\textwidth
\epsfbox{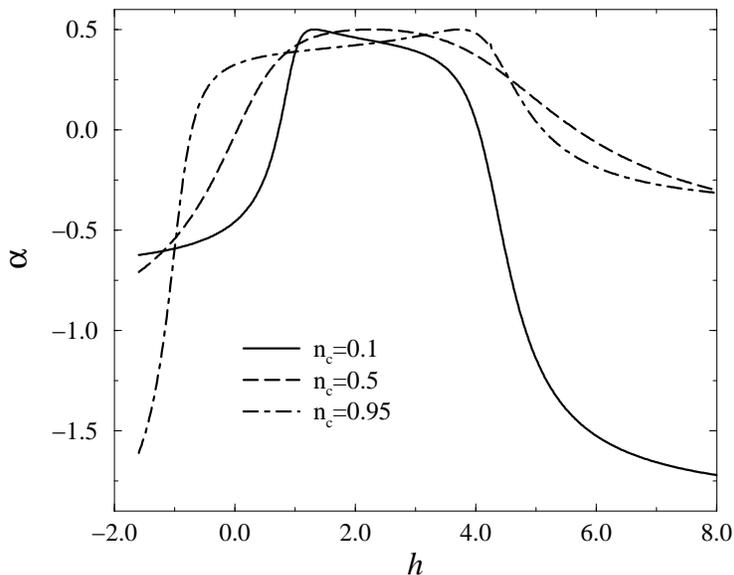}
\end{center}
\caption{\label{fig:hubb_a}%
Exponents at the {\em absolute} threshold for X-ray absorption in the
Hubbard model as a function of the boundary chemical potential $h$ for
$u=1$ and several densities $n_c$.
}
\end{figure}

Since we consider the Hubbard model at less than half
filling (i.e.\ $n_c<1$) this implies that a positive exponent $\alpha$
leading to a edge singularity is possible {\em only} in the intermediate
regime.  The corresponding numerical data for finite $u$ are presented in
Figure~\ref{fig:hubb_a}.

\subsection{Intermediate thresholds}
\label{sec:hinter}
Finally we consider some cases where the absorption excites the system into
a state in which some but not all bound states are occupied.  First,
let the final state be characterized by {\em one} antiholon and {\em
one} spinon in a bound state which gives rise to a singularity at an
energy {\em between} the two thresholds discussed above. 
Such a process is possible for boundary potentials $h>u+\sqrt{u^2+1}$ (or
$S>u$) and corresponds to a Bethe Ansatz state with {\em a single} complex
$k$.  Analyzing the Bethe Ansatz equations we obtain the relation
$\theta^s={1\over2}\theta^c$.  In this case $\theta^c$ has to be computed
from Eqs.~(\ref{thc:h}) and (\ref{irho:h}) with ${\bar g}_c(k)$ given by
(\ref{gc1:h}) with
\begin{equation}
    f_b(x) = -a_{2S}(x) 
      + {1\over u}G_{{S\over u}+1}\left({x\over2u}\right) .
\end{equation}
The finite size spectrum is again of the form (\ref{fsop}); the quantum
numbers of the relevant final state are $\Delta N_c^0=0=\Delta N_s^0$.
{}From (\ref{exp}) we obtain
\begin{equation}
  {\alpha}_{int}^{} = {3\over4} - {1\over2\xi^2}
  \left(\theta^c\right)^2
\end{equation}
for the edge exponent determining the singularity at this threshold.  In
the strong coupling limit we find that ${\alpha}_{int}^{}$ varies
between $5/8$ for the empty band and $-3/8$ as we approach half-filling.
An edge singularity can be observed for $n_c<\sqrt{{3\over2}} -
{1\over2}\approx 0.725$.

A different intermediate thershold occurs if {\em only} an antiholon
is in one of the bound states.  This final state is already possible for
$h>1$ and is parametrized by a single complex root $k=i\ln h$ for $S<u$ and
an additional complex spin rapidity $\lambda$ for $S>u$.  Depending on $h$
several cases have to be distinguished resulting in a edge singularity with
exponent
\begin{equation}
  {\alpha}_{int}^{\prime} = {1\over2} - {1\over2\xi^2}
  \left(\theta^c\right)^2\
\end{equation}
for $S<3u$ (for $S>3u$ the exponent is always negative).  The function
$f_b(x)$ in (\ref{gc1:h}) is now simply one half of that in (\ref{fb1:h}).
In the strong coupling limit the edge exponent ${\alpha}_{int}^{\prime}$
can be expressed through $n_c$ and $h$ using Eq.~(\ref{thinf0}).  In this
limit a singularity in the absorption spectrum (i.e.\ positive exponent)
can be observed for sufficiently large boundary potentials $h\gtrsim
\tan\left({\pi\over4}(2n_c+1)\right)>1$ as long as $n_c<{1\over2}$ but
only close to $h\approx4u$ above quarter filling. 

Note that for sufficiently strong boundary potentials the cases discussed
here are only a small subset of the possible thresholds. Furthermore, for
sufficiently strong {\em repulsive} boundary potentials, i.e. $h<-1$, the
spectrum allows for holon bound states. Like in the case of the
$t$-$J$ model with attractive boundary chemical potentials there
exist solutions to the Bethe equations with complex quasi momenta
$k=\pi+i\ln|h|$ of Eq.~(\ref{bae:h}) that have a gap with respect to
the absolute ground state and lead to a higher threshold in the
X-ray spectrum. 
\section{Conclusions}
In this work we have determined the X-ray edge exponents in a Luttinger
liquid for the case where the local disturbance due to the core hole leads
to bound states.  We used specific realizations of Luttinger liquids on the
lattice, namely Hubbard and $t$-$J$ models with integrable boundary terms.
The main difference to the Fermi liquid case (\ref{hfermi}) solved in
Refs.~\onlinecite{cono,affl:96} is that due to spin and charge separation
we find a richer structure of thresholds in the X-ray absorption rate
associated with bound states of spinons and (anti)holons. Using Boundary
Conformal Field Theory the exact dependence of the edge exponents on
band filling and interaction strength can be extracted from the finite size
spectra which are determined from the Bethe Ansatz solution.

For weak boundary fields our results coincide with those obtained in a
field theoretical treatment by Prokof{'}ev\cite{prok:94} and Affleck and
Ludwig\cite{aflu:94} if the boundary chemical potentials are fine-tuned.

For sufficiently strong boundary fields the models considered in this paper
allow for various bound states, each of which can lead --- in principle ---
to a singularity in the absorption spectrum.  Previous studies of
these additional singularities have not taken into account the interaction
between the particles in the bound states and those remaining in the band
\cite{cono,affl:96}. This results in a simple relation between the exponents
at different edges with the phase shift $\delta(\epsilon_F)$ at the Fermi
surface as the only free parameter. In the systems considered here
the occupation of the boundary bound states modifies the potential
acting on the particles remaining in the bands which in turn modifies
the corresponding phase shifts.  Examining the edge exponents for the
different thresholds we find that for generic values of boundary
potentials and filling factors many of them will in fact be negative,
and consequently won't lead to an observable singularity in the spectrum.

\section*{Acknowledgements}
We thank Y. Avishai, J. Chalker, A. Jerez, T. Kopp D. Lee and
A.M. Tsvelik for helpful discussions. FHLE is supported by the EU
under Human Capital and Mobility fellowship grant ERBCHBGCT940709. HF
is supported in parts by the Deutsche For\-schungs\-gemeinschaft under
Grant No.\ Fr~737/2--2.

\end{document}